\documentclass[letterpaper,10pt,conference]{ieeeconf}

\usepackage{cite}
\usepackage{graphicx}
\usepackage{subcaption}
\usepackage[cmex10]{amsmath}
\usepackage{algpseudocode}
\usepackage{algorithmicx}
\usepackage{xcolor}
\usepackage{array}
\usepackage{url}
\usepackage{bbm}
\usepackage{todonotes}

\usepackage{epsfig}
\usepackage{amsfonts,amssymb,multirow,bigstrut,booktabs,ctable,latexsym}
\usepackage{mathtools}
\usepackage[mathscr]{eucal}
\usepackage{verbatim}

\usepackage{amsthm}
\usepackage{algorithm}
\usepackage[normalem]{ulem}
\newtheorem{definition}{Definition}
\newtheorem{lemma}{Lemma}
\newtheorem{theorem}{Theorem}

\newtheorem{proposition}{Proposition}
\newtheorem{remark}{Remark}
\newtheorem{assumption}{Assumption}

\newtheorem{problem}{Problem}

\IEEEoverridecommandlockouts
	\title{\bf Privacy-Preserving Resilience of Cyber-Physical Systems to Adversaries}
	
	\author{Bhaskar Ramasubramanian$^{1}$, Luyao Niu$^{2}$, Andrew Clark$^{2}$, Linda Bushnell$^{1}$, and Radha Poovendran$^{1}$%
		\thanks{$^{1}$Network Security Lab, Department of Electrical and Computer Engineering, 
			University of Washington, Seattle, WA 98195, USA. \newline
			{\tt\small \{bhaskarr, lb2, rp3\}@uw.edu}}
\thanks{$^2$Department of Electrical and Computer Engineering, Worcester Polytechnic Institute, Worcester, MA 01609 USA.
			{\tt\{lniu,aclark\}@wpi.edu}}
		\thanks{This work was supported by the U.S. Army Research Office, National Science Foundation, and the Office of Naval Research via Grants W911NF-16-1-0485, CNS-1941670, and N00014-17-S-B001 respectively. }
	}
\begin{document}	
	\maketitle
	
	
\begin{abstract}
A cyber-physical system (CPS) is expected to be resilient to more than one type of adversary. 
In this paper, we consider a CPS that has to satisfy a linear temporal logic (LTL) objective in the presence of two kinds of adversaries. 
The first adversary has the ability to tamper with inputs to the CPS to influence satisfaction of the LTL objective. 
The interaction of the CPS with this adversary is modeled as a stochastic game. 
We synthesize a controller for the CPS to maximize the probability of satisfying the LTL objective under any policy of this adversary. 
The second adversary is an eavesdropper who can observe labeled trajectories of the CPS generated from the previous step. It could then use this information to launch other kinds of attacks. 
A labeled trajectory is a sequence of labels, where a label is associated to a state and is linked to the satisfaction of the LTL objective at that state. 
We use differential privacy to quantify the indistinguishability between states that are related to each other when the eavesdropper sees a labeled trajectory. 
Two trajectories of equal length will be differentially private if they are differentially private at each state along the respective trajectories. 
We use a skewed Kantorovich metric to compute distances between probability distributions over states resulting from actions chosen according to policies from related states in order to quantify differential privacy. Moreover, we do this in a manner that does not affect the satisfaction probability of the LTL objective. 
We validate our approach on a simulation of a UAV that has to satisfy an LTL objective in an adversarial environment.
\end{abstract}

\section{Introduction}\label{Intro}
	
Cyber-physical systems (CPSs) consist of tightly coupled cyber and physical components that work in conjunction with algorithms and communication channels to satisfy complex objectives in dynamic environments \cite{baheti2011cyber}. 
The objectives that a CPS will need to meet often vary with time. 
Temporal logic frameworks like linear temporal logic \cite{baier2008principles} enable the specification of goals such as safety, stability, and reachability. 
In applications like power systems \cite{sullivan2017cyber} and automobiles \cite{shoukry2013non}, the CPS must achieve its goals in scenarios that can be manipulated by an adversary that can disrupt nominal operation \cite{banerjee2012ensuring}. 
Disruptions to power systems and water networks can inconvenience large sections of the population. 

An adversary may not have the ability to directly launch attacks on the CPS. 
However, it could be capable of collecting information about the system, which can then be used to affect nominal operation \cite{focardi1994taxonomy}. 
Therefore, a well-designed system must protect information critical to maintaining normal operation. 
Differential privacy \cite{dwork2008differential} is a security property that makes it difficult for an adversary to discern information about a system by providing probabilistic guarantees on the indistinguishabilty of observations of the system. 
This technique was originally used to protect sensitive data of individuals in databases in a manner that allowed for statistical analysis on aggregated information obtained from the data \cite{dwork2006calibrating}. 
It has since been used to protect information represented as trajectories of a dynamical system \cite{le2013differentially}, where differential privacy was accomplished by adding a carefully calibrated noise to sensitive trajectories so that an adversary could not glean information about the modified trajectory. 
Recent work has studied the enforcement of differential privacy on multi-agent systems to meet a temporal logic objective \cite{xu2019differentially}, or to minimize a quadratic cost \cite{hale2018privacy}.

We study differential privacy for systems represented as discrete state, discrete-action Markov decision processes (MDPs). 
Such systems encompass the operating environments of a broad range of practical systems, including robots and UAVs \cite{kress2007s}, and are especially suited for experimental analysis.  
In order to reason about trajectories that satisfy an LTL formula, we work with $PCTL^*$, a probabilistic computational tree logic which has LTL formulas as its path formulas \cite{baier2008principles}. 
Two states will be differentially private if the probabilities of satisfying a desired LTL objective starting from these states are sufficiently close. 
A trajectory will be differentially private if all states in the trajectory that are generated according to some policy are differentially private. 
In previous work that studied differential privacy for Markov chains \cite{chistikov2018bisimilarity, chistikov2019asymmetric}, the treatment was restricted to ensuring differential privacy of the initial state of the Markov chain. 
In comparison, we study differential privacy of states in an MDP along trajectories corresponding to satisfaction of an LTL objective in the presence of an adversary. 

Different from the aforementioned works, we consider a setting with two kinds of adversaries that have different capabilities, and act independently of each other. 
We are interested in ensuring differential privacy of trajectories against one adversary ($E$). 
At the same time, we want to maximize the probability of satisfaction of the LTL objective under actions of the other adversary ($A$). 
Specifically, we assume that adversary $A$ has the ability to inject signals to affect the control inputs to the CPS thereby influencing the transitions between CPS states. 
We would like to maximize the probability of satisfying the LTL goal under any sequence of actions played by this adversary. 
The interaction between the CPS and $A$ is modeled as a zero-sum game, and the synthesis of a CPS policy that maximizes the probability of satisfying the LTL goal under any adversary policy is related to reaching a Stackelberg equilibrium of this game. 
Once this objective is accomplished, we want that trajectories produced by the synthesized controller be indistinguishable to the adversary $E$ that can eavesdrop on these trajectories by observing a sequence of labels associated to states along the trajectories. 
That is, observing labeled trajectories should not allow $E$ to gain information about the state of the system. 
To the best of our knowledge, this is the first work that studies resilient control in the presence of an adversary along with ensuring differential privacy of states along a trajectory satisfying the LTL objective. 
Prior work has studied these problems separately for a single kind of adversary (either an adversary of type $A$ or of type $E$).

\subsection{Contributions}

We solve the problem of satisfying an LTL objective in the presence of an adversary that can tamper with actuator inputs, while also ensuring that trajectories of the system which satisfy the objective are differentially private to an eavesdropper who can observe labels of states on these trajectories. 
We make the following contributions:

\begin{itemize}

\item We show that the 
CPS policy that will maximize the probability of satisfaction of the LTL objective under any adversary policy is related to reaching certain subsets of a Markov chain associated to representations of the environment and the LTL goal under these policies. 

\item We use a fragment of the logic $PCTL^*$, whose semantics are defined over MDPs, to reason about LTL trajectories. Starting with a realization of the defender policy that satisfies the LTL objective, we define a symmetric relation on states and show that pairs of states in this relation will be differentially private. 

\item We present a value-iteration procedure and show that if a (skewed) distance between values of `related' states is below a threshold, then these states will be differentially private. 
Two trajectories will be differentially private if the initial states of the trajectories are related, and if optimal defender policies from these states and all pairs of subsequently related states are `close' to each other. 

\item A case-study on the satisfaction of a reach-avoid specification for a UAV influenced by adversarial inputs in an environment with an eavesdropper illustrates our results. 
\end{itemize}

\subsection{Outline of Paper}

The rest of this paper is organized as follows: 
Section \ref{RelatedWork} summarizes related work, and we give a brief introduction to LTL, stochastic games, and differential privacy in Section \ref{Preliminaries}. 
We state our problem and detail the steps of our solution in Sections \ref{Problem} and \ref{SolutionApproach}. 
Section \ref{Example} shows an example illustrating our approach, and Section \ref{Conclusion} concludes the paper. 
	
\section{Related Work}\label{RelatedWork}

The satisfaction of an LTL objective for two-player stochastic games when the players had competing objectives was presented in \cite{niu2018secure, niu2019optimal} for the case when states were fully observable and in \cite{ramasubramanian2019secure, bhaskar2019secure} for the partially observable case. 
The authors of \cite{niu2020control, niu2020robust} extended this to the case when an adversary could tamper with both actuators and clocks of the CPS to affect the satisfaction of a time-sensitive temporal constraint. 
There, the same adversary was capable of effecting both kinds of attacks on the system, which makes it different from the assumptions made in this paper. 
Evaluating multiple traces corresponding to different executions of a system using hyperproperties \cite{clarkson2010hyperproperties} was proposed in \cite{abraham2018hyperpctl}. 
The authors proposed a temporal logic $HyperPCTL$, in order to reason about probabilistic hyperproperties, and this approach was studied in the context of verification for CPSs in \cite{wang2019statistical}. 

The authors of \cite{yang2017differential, castiglioni2020logical} presented a logical characterization of differential privacy for labeled probabilistic transition systems using a trace metric that corresponded to \emph{exact} differential privacy ($\delta = 0$ in Definition \ref{DiffPrivDefn}). 
The authors of \cite{chistikov2018bisimilarity, chistikov2019asymmetric} computed distances that constituted sound upper bounds on a skewed total variation distance between initial states of a Markov chain. 
They used the Kantorovich metric \cite{kantorovich1942transfer} 
to lift a distance between states to a distance between probability distributions associated to these states. 
A large part of this analysis was motivated by the work on defining metrics for MDPs \cite{desharnais2004metrics, ferns2004metrics, ferns2011bisimulation}. 
These metrics measured the distance between states in MDPs with large state spaces and then allowed aggregation of states that were `close' to each other. 
They further showed that for such states, the values of the states were also `close' to each other. 
Algorithms based on Monte-Carlo methods were used to give guarantees on the differential privacy of synthesized policies in \cite{balle2016differentially}. 

Using differential privacy to protect information about trajectories of dynamical systems was first presented in \cite{le2013differentially}. 
Since then, it has been studied for privacy-preserving consensus in multi-agent dynamical systems \cite{mo2016privacy}, networked systems \cite{cortes2016differential}, and linear quadratic control for multi-agent systems \cite{hale2018privacy}. 
We point the reader to the review in \cite{han2018privacy}, and references therein for a detailed survey of recent developments using differential privacy in dynamical systems. 

Differential privacy in the context of the satisfaction of temporal objectives is a relatively recent area of research. 
In \cite{liu2018model}, the authors introduced $dpCTL^*$, an extension of the logic $PCTL^*$ augmented with a differentially private operator. 
They presented a model-checking procedure to verify differential privacy on Markov chains. 
The authors assumed that actions are controlled by an adversary, which is different from the setup in this paper. 
The authors of \cite{xu2019differentially} proposed a differentially private controller synthesis procedure for multi-agent systems to satisfy a metric temporal logic objective. 
In their setting, each agent added a noise term while communicating its location to a local hub, which in turn transmitted the information to a cloud controller that determined the optimal inputs for each agent. 


\section{Preliminaries}\label{Preliminaries}

\subsection{Linear Temporal Logic}\label{LTL}

Temporal logic frameworks enable the representation and reasoning about temporal information on propositional statements. 
\emph{Linear temporal logic (LTL)} is one such framework, where the progress of time is `linear'. 
An \emph{LTL formula} \cite{baier2008principles} is defined over a set of atomic propositions $\mathcal{AP}$, and can be written as $\varphi:=\mathtt{T}|\sigma| \neg \varphi | \varphi \wedge \varphi | \mathbf{X} \varphi |\varphi \mathbf{U} \varphi$, 
where $\sigma \in \mathcal{AP}$, and $\mathbf{X}$ and $\mathbf{U}$ are temporal operators denoting the \emph{next} and \emph{until} operations respectively. 
The semantics of LTL are defined over (infinite) words in $2^{\mathcal{AP}}$. 
We write $\eta_0 \eta_1\dots:=\eta \models \varphi$ when a trace 
$\eta \in (2^{\mathcal{AP}})^{\omega}$ 
satisfies an LTL formula $\varphi$. 
%
\begin{definition}[LTL Semantics]\label{LTLSemantics}
Let $\eta^i = \eta_i\eta_{i+1}\dots$. 
Then, the semantics of LTL can be recursively defined as: 
\begin{enumerate}
\item $\eta \models \mathtt{T}$ if and only if (iff) $\eta_0$ is true; 
\item $\eta \models \sigma$ iff $\sigma \in \eta_0$; 
\item $\eta \models \neg \varphi$ iff $\eta \not \models \varphi$; 
\item $\eta \models \varphi_1 \wedge \varphi_2$ iff $\eta \models \varphi_1$ and $\eta \models \varphi_2$; 
\item $\eta \models \mathbf{X} \varphi$ iff $\eta^1 \models \varphi$; 
\item $\eta \models \varphi_1 \mathbf{U} \varphi_2$ iff $\exists j \geq 0$ such that $\eta^j \models \varphi_2$ and for all $k < j, \eta^k \models \varphi_1$. 
\end{enumerate}
\end{definition}

Moreover, the logic admits derived formulas of the form: 
\emph{i)} $\varphi_1 \vee \varphi_2:=\neg(\neg \varphi_1 \wedge \neg \varphi_2)$; 
\emph{ii)} $\varphi_1 \Rightarrow \varphi_2:= \neg \varphi_1 \vee \varphi_2$; 
\emph{iii)} $\mathbf{F}\varphi:=\mathtt{T} \mathbf{U} \varphi  \text{  (eventually)}$; 
\emph{iv)} $\mathbf{G} \varphi:= \neg \mathbf{F} \neg \varphi \text{  (always)}$. 
\begin{definition}[Deterministic Rabin Automaton]
A \emph{deterministic Rabin automaton (DRA)} is a quintuple $\mathcal{RA} = (Q, \Sigma, \kappa, q_0,F)$ where $Q$ is a nonempty finite set of states, $\Sigma$ is a finite alphabet, $\kappa : Q \times \Sigma \rightarrow Q$ is a transition function, $q_0 \in Q$ is the initial state, and $F:=\{(L(i),K(i)\}_{i=1}^M$ where $L(i), K(i) \subseteq Q$ for all $i$, and $M$ is a positive integer. 
\end{definition}

A \emph{run} of $\mathcal{RA}$ is a sequence of states $q_0q_1\dots$ such that $q_{i} \in \kappa (q_{i-1}, \alpha)$ for all $i$ and for some $\alpha \in \Sigma$. 
The run is \emph{accepting} if for some $(L,K) \in F$, the run intersects with $L$ finitely many times, and with $K$ infinitely often. 
An LTL formula $\varphi$ over $\mathcal{AP}$ can be represented by a DRA with alphabet $2^{\mathcal{AP}}$  that accepts all and only those runs that satisfy $\varphi$. 
%
\subsection{Labeled Stochastic Games and Markov Chains}\label{SGMC}

A stochastic game (SG) involves two players, and starts with the system in a particular state. 
Transitions to subsequent states are probabilistically determined by the current state and the actions chosen by each player.
%
\begin{definition}[Stochastic Game]\label{SGDefn}
A \emph{stochastic game} \cite{niu2018secure} is a tuple $\mathcal{G}:=(S, U_{def}, U_{adv}, \mathbb{T}, \mathcal{AP}, \mathcal{L})$. 
$S$ is a finite set of states, $U_{def}$ and $U_{adv}$ are finite sets of actions of the defender and adversary. 
The function $\mathbb{T}: S \times U_{def} \times U_{adv}  \times S \rightarrow [0,1]$ encodes $\mathbb{T}(s'|s,u_{def},u_{adv})$, the probability of transition from state $s$ to state $s'$ when defender and adversary actions are $u_{def}$ and $u_{adv}$. 
$\mathcal{AP}$ is a set of atomic propositions. $\mathcal{L}: S \rightarrow 2^{\mathcal{AP}}$ is a labeling function that maps a state to a subset of atomic propositions that are satisfied in that state. 
\end{definition}

SGs can be viewed as an extension of Markov Decision Processes (MDPs) when there is more than one player taking an action. 
For a player in an SG, a \emph{policy} is a mapping from sequences of states to actions, if it is deterministic, or from sequences of states to a probability distribution over actions, if it is randomized. 
A policy is \emph{stationary} if it is dependent only on the most recent state. 
We denote the defender's policy by $\mu$ and the adversary's policy by $\tau$. 

In this paper, we focus our attention on the Stackelberg setting \cite{fudenberg1991game}, where the first player (leader) commits to a policy. 
The second player (follower) observes this and chooses its policy as the best response to the leader's policy, defined as the policy that maximizes the follower's utility. 
We assume that the players take their actions concurrently at each time step. 
%
We define the notion of a Stackelberg equilibrium (SE), which indicates that a solution to a Stackelberg game has been found. 
Let $Q_L(\mathit{l}, \mathit{f})$ ($Q_F(\mathit{l}, \mathit{f})$) be the utility gained by the leader (follower) by adopting a policy $\mathit{l}$ ($\mathit{f}$). 
\begin{definition}[Stackelberg Equilibrium]\label{StackEq}
A pair $(\mathit{l}, \mathit{f})$ is a \emph{Stackelberg equilibrium} if $\mathit{l} = \arg \max_{\mathit{l}'}Q_L(\mathit{l}', BR(\mathit{l}'))$, where $BR(\mathit{l}') = \{\mathit{f}: \mathit{f}=\arg \max Q_F(\mathit{l}',\mathit{f})\}$. 
That is, the leader's policy is optimal given that the follower observes the leader's policy and plays its best response. 
\end{definition}

Given $\mu$ and $\tau$, $\mathcal{G}$ is a \emph{Markov chain (MC)} \cite{meyn2012markov}. 
For $s, s' \in S$, $s'$ is \emph{accessible} from $s$, written $s \rightarrow s'$, if 
$\mathbb{P}(s_a|s) \mathbb{P}(s_b|s_a) \dots \mathbb{P}(s_i|s_j) \mathbb{P}(s'|s_i) > 0$ for some (finite subset of) states $s_a,s_b,\dots,s_i,s_j$. 
Two states \emph{communicate} 
if $s \rightarrow s'$ and $s' \rightarrow s$. 
\emph{Communicating classes} of states cover the state space of the MC. 
A state is \emph{transient} if there is a nonzero probability of not returning to it when we start from that state, and is \emph{positive recurrent} otherwise. 
In a finite state MC, every state is either transient or positive recurrent. 
%
%
We formally define labeled (discrete-time) Markov chains, and a measure on this entity below.
\begin{definition}[Labeled Markov Chain]\label{LabeledMarkovChain}
A labeled Markov chain is a tuple $\mathcal{M} = (S, \mathbb{T}, \mathcal{AP}, \mathcal{L})$, where $S,\mathcal{AP}, \mathcal{L}$ are as in Definition \ref{SGDefn}, and $\mathbb{T}: S \times S \rightarrow [0,1]$ encodes $\mathbb{P}(s'|s)$\footnote{We further assume that exactly one atomic proposition will be true in a state of $\mathcal{M}$. Therefore, $\mathcal{L}:S \rightarrow \mathcal{AP}$. This is not restrictive since if labels $l_1, l_2$ are true, we can define $l_{12} = l_1 \wedge l_2$, and augment $l_{12}$ to $\mathcal{L}$.}. 
A path on $\mathcal{M}$ is a sequence of states $\mathit{S} = s_0s_1,\dots$ such that $s_i \in S$ and $\mathbb{T}(s_{i+1},s_i) > 0$ for all $i$. 
We write $\mathtt{Paths}(s)$ to denote the set of paths in $\mathcal{M}$ that start from state $s$. 
\end{definition}

A labeled MDP is defined by augmenting a finite set of actions $Act$ to the Markov chain in Definition \ref{LabeledMarkovChain}, and redefining the transition function as $\mathbb{T}: S \times Act \times S \rightarrow [0,1]$ to encode $\mathbb{P}(s'|s, a)$. 
A policy on this labeled MDP is a map from sequences of states to actions (or, distributions over actions). 
In order to distinguish between policies synthesized to satisfy the LTL specification, and policies designed to ensure differential privacy, we will refer to the latter as a \emph{scheduler}, denoted $\Gamma$, and the Markov chain so induced is $\mathcal{M}^{\Gamma}$. 
In this paper, we will assume that the output of the scheduler will depend only on the most recent state. 

To quantitatively reason about $\mathcal{M}$, we need to define an appropriate probability space. 
We follow the treatment in \cite{baier2008principles}, and let the sample space be $\mathtt{Paths}(s)$. 
The set of events $\mathcal{F}$ is the smallest $\sigma-$algebra on $\mathtt{Paths}(s)$ that contains all \emph{cylinder sets}\footnote{The cylinder set of $\omega = s_0s_1\dots,s_n$ is $Cyl(\omega):=\{\omega ' \in \bigcup_s \mathtt{Paths}(s)| \omega \text{ is a prefix of } \omega'\}$.} spanned by finite length paths in $\mathcal{M}$. 
Then, there is a unique probability measure on $\mathcal{M}$ associated to this $\sigma-$algebra. 
However, this measure is a value on the cylinder sets. 
The next definition assigns a measure to an arbitrary state of $\mathcal{M}$ \cite{chistikov2018bisimilarity, chistikov2019asymmetric}. 

\begin{definition}[Measure on $s \in S$ in $\mathcal{M}$]
Let $\mathbb{T}^+(s_0,s_1,\dots,s_k)$ $ =  \prod_{i=0}^{k-1}\mathbb{P}(s_{i+1}|s_i)$ and $\mathcal{L}^+(s_0,\dots,s_k)$ $ = \mathcal{L}(s_0)\dots\mathcal{L}(s_k)$. 
Then, for $s \in S$, $\nu_s:\mathcal{F} \rightarrow [0,1]$ is the unique measure on $\mathcal{F}$ such that for any cylinder set $Cyl(\omega)$, $\nu_s(Cyl(\omega)) = \sum \{\mathbb{T}^+(p): p \in \mathtt{Paths}(s), \mathcal{L}^+(p) = \omega\}$
\end{definition}

\subsection{Differential Privacy}\label{DiffPrivLMC}

Differential privacy is a property that ensures that private data of an agent is protected, while allowing for statistical inferences from aggregates of the data \cite{dwork2014algorithmic}. 
This makes it unlikely that an adversary will learn anything meaningful about sensitive data. 
Attractive features of differential privacy include compositionality, resilience to post-processing, and robustness to side information. 
The notion of differential privacy that we use in this paper is defined over $\mathcal{M}$. 

\begin{definition}[$(\epsilon, \delta)-$Differential Privacy]\label{DiffPrivDefn}
Let $R \subset S \times S$ be a symmetric relation. 
Then, given $\epsilon \geq 0$ and $\delta \in [0,1]$, a labeled Markov chain $\mathcal{M}$ is $(\epsilon, \delta)-$differentially private with respect to $R$ if for every $s, s' \in S$ such that $(s, s') \in R$, $\nu_s(E) \leq e^{\epsilon} \nu_{s'}(E) + \delta$ for every measurable subset $E \in \mathcal{F}$. 
\end{definition} 

A measure of how different two states $s, s' \in S$ in $\mathcal{M}$ can be quantified using 
the \emph{total variation distance}, given by:
\begin{align*}
tv(s,s') := tv(\nu_s,\nu_{s'})= \sup_{E \in \mathcal{F}} |\nu_s(E) - \nu_{s'}(E)|
\end{align*}
	
\section{Problem Formulation}\label{Problem}

\begin{problem}\label{ProblemOrig}
Given a stochastic game $\mathcal{G}$ representing the environment, an LTL specification $\varphi$, and parameters $\epsilon, \delta$: \emph{i)}: determine a defender policy $\mu$ to maximize the satisfaction probability of $\varphi$ under any adversary policy $\tau$; 
\emph{ii)}: determine a symmetric relation $R$ so that for any trajectory from \emph{i)} that satisfies $\varphi$, there is some other trajectory such that states $s, s'$ along the two trajectories are in $R$, and the two trajectories are $(\epsilon, \delta)-$differentially private to an eavesdropper.

%
\end{problem}

We consider two adversaries, each having different capabilities. 
The first adversary can inject inputs into the system in order to influence transitions between states in the CPS. 
This sequence of inputs of this adversary is determined by the policy $\tau$ in Problem \ref{ProblemOrig}. 
The second adversary is an eavesdropper, who can observe trajectories of the system as a result of the policy synthesized by the defender, and can potentially use this information to launch an attack on the system. 
Ensuring differential privacy of this aforementioned trajectory will ensure that states along the trajectory are relatively indistinguishable to this eavesdropper. 
\begin{assumption}
The two adversaries act independent of each other, and do not communicate with each other.
\end{assumption}

\section{Solution Approach} \label{SolutionApproach}

We adopt a two-step approach to solve Problem \ref{ProblemOrig}. 
In the first step, we will determine a defender policy to maximize the probability of satisfying the LTL objective under any adversary policy. 
In the second step, we will use optimal policies that satisfy $\varphi$ from the previous step to define a symmetric relation $R$ on the states 
to ensure $(\epsilon, \delta)-$differential privacy of a labeled trajectory observed by the eavesdropper. 
We will impose an additional constraint on the `closeness' of policies from states that are `related' to each other (this will be made precise later in this section) to ensure differential privacy of subsequent states along the trajectory while maintaining satisfaction of the LTL objective. 
%

The representation of the environment when composed with that of $\varphi$ yields a product game. 
For an instantiation of defender and adversary policies, this is a Markov chain. 
These policies will induce paths on the Markov chain, and we want to ensure differential privacy of states along these paths. 
We will use a fragment of the temporal logic PCTL* \cite{baier2008principles} to reason about probabilities over paths on this entity. 

\subsection{Synthesis of Policies to Satisfy LTL Objective}\label{SynthDefPolicy}

In order to find runs on $\mathcal{G}$ that would be accepted by the DRA $\mathcal{RA}$ corresponding to the LTL task $\varphi$, we construct an entity that composes representations of the environment ($\mathcal{G}$) and the goal ($\mathcal{RA}$). 
We call this a \emph{product game}. 

\begin{definition}[Product Stochastic Game (PSG)]
Given SG $\mathcal{G}$ and DRA $\mathcal{RA}$ corresponding to LTL formula $\varphi$, a PSG is a tuple $\mathcal{G}^{\varphi} :=(S^{\varphi}, U_{def}, U_{adv}, \mathbb{T}^{\varphi}, F^{\varphi}, \mathcal{AP}, \mathcal{L}^{\varphi})$, where $S^{\varphi} = S \times Q$, $\mathbb{T}^{\varphi} ((s',q')|(s,q),u_{def},u_{adv}) = \mathbb{T}(s'|s,u_{def},u_{adv})$ iff $\delta(q, \mathcal{L}(s')) = q'$ and $0$ otherwise, $F^{\varphi}:=\{(L^{\varphi}(i),K^{\varphi}(i)\}_{i=1}^M$ is such that $L^{\varphi}(i), K^{\varphi}(i) \subseteq S^{\varphi}$, and $(s,q) \in L^{\varphi}(i)$ iff $q \in L(i)$ and $(s,q) \in K^{\varphi}(i)$ iff $q \in K(i)$, $\mathcal{L}^{\varphi}((s,q)) = \mathcal{L}(s)$.
\end{definition}

As a first step, we are interested in the synthesis of defender policies that would satisfy the LTL objective under any adversary policy. 
This will be equivalent to reaching certain recurrent subsets of a Markov chain formed under instantiations of these policies. 
Maximizing the probability of satisfaction in this setting corresponds to reaching an equilibrium of a zero-sum Stackelberg game between the defender and adversary. 
A careful justification of this assertion with detailed proofs for fully and partially observable environments has been studied in our earlier works \cite{niu2018secure, niu2019optimal, ramasubramanian2019secure, bhaskar2019secure}. 
We only state relevant results that will be useful in our goal to further establish differential privacy of trajectories in $\mathcal{G}^{\varphi}$ that will satisfy $\varphi$ under the respective agent policies. 

\begin{proposition}
Let $v(s,q) = \max \limits_\mu \min \limits_\tau \mathbb{P}(\varphi|\mu, \tau, (s,q))$. 
Then, 
\begin{align*}
v(s,q)&=\max_\mu \min_\tau \sum_{u_d \in U_{def}} \sum_{u_a \in U_{adv}} \sum_{(s',q') \in S^\varphi}\mu(u_{def}|(s,q))  \\\times ~&\tau (u_{adv}|(s,q)) \mathbb{T}^{\varphi} ((s',q')|(s,q),u_{def},u_{adv})v((s',q')) 
\end{align*}
\end{proposition}
\begin{proof}
The proof can be found in Lemma 1 of \cite{niu2019optimal}.
\end{proof}

Let $\mathcal{E}$ denote the set of \emph{accepting states} of $\mathcal{G}^{\varphi}$. 
That is, a subset of recurrent states of the stochastic game which also satisfy $\varphi$. 
We note that in the terminology of \cite{niu2018secure, niu2019optimal} this set is part of a \emph{generalized maximal accepting end component}, while \cite{ramasubramanian2019secure, bhaskar2019secure} use the term \emph{$\varphi-$ feasible recurrent set}. 
Let $\mathbb{P}(\texttt{reach} ~ \mathcal{E}|s, \mu, \tau)$ denote the probability of reaching the set of states $\mathcal{E}$ in $\mathcal{G}^{\varphi}$. We have the following result for stationary policies $\mu, \tau$. 

\begin{theorem}
For a stationary defender policy $\mu$, and initial state $s$, the following holds: 
\begin{align*}
\min_{\tau} \mathbb{P}(\mathcal{G}^{\varphi} \models \varphi |s, \mu, \tau)&= \min_\tau \mathbb{P}(\texttt{reach} ~ \mathcal{E}|s, \mu, \tau)
\end{align*}
\end{theorem}
\begin{proof}
The proof can be found in Proposition 3 of \cite{niu2019optimal}. An analogous result for the partially observable state setting can be found in Theorem 4.3 of \cite{ramasubramanian2019secure}.
\end{proof}

Therefore, the problem of maximizing the probability of satisfaction of $\varphi$ under any adversary policy is equivalent to reaching a subset of states under these policies of the product game $\mathcal{G}^{\varphi}$ that composes representations of the environment and the LTL objective. 
Moreover, results in \cite{niu2018secure, niu2019optimal, ramasubramanian2019secure, bhaskar2019secure} establish convergence of these policies to an equilibrium of the Stackelberg game between the defender and adversary. 

\subsection{The Temporal Logic PCTL*}

The semantics of LTL, as seen in Definition \ref{LTLSemantics} are defined over infinite words. 
In order to establish differential privacy, we require a means to reason over paths in a Markov chain or Markov decision process. 
This temporal logic PCTL* is appropriate in this setting. 
A PCTL* formula comprises state and path formulas and its semantics are expressed over Markov chains or MDPs. 
We will work with a fragment of PCTL* whose path formulas are LTL formulas. 
	
An \emph{PCTL* state formula} \cite{baier2008principles} is defined over a set of atomic propositions $\mathcal{AP}$, and can be written as: $\Phi:=\mathtt{T}|\sigma| \neg \Phi | \Phi \wedge \Phi | \mathbb{P}_J(\phi)$, 
where $\phi$ is a \emph{path formula}, which is an LTL formula whose sub-state formulas are PCTL* state formulas. 
This is formed according to $\phi:=\Phi| \neg \phi | \phi \wedge \phi | \mathbf{X} \phi |\phi \mathbf{U} \phi$,  
where $\Phi$ is a PCTL* state formula. $J \subseteq [0,1]$ is a non-empty interval with rational end points. 
%
The semantics of PCTL* are expressed over an MDP $\mathcal{M}^{\Gamma}$, where $\Gamma$ is a scheduler. 
In the sequel, we omit the superscript $\Gamma$, and use it only when the context will not be clear otherwise. 

\begin{definition}[PCTL* Semantics]\label{PCTL*Semantics}
The semantics of PCTL* state formula are defined on states of a labeled MDP $\mathcal{M}$ as:
\begin{enumerate}
\item $(\mathcal{M},s) \models \mathtt{T}$ iff $\mathtt{Paths}(s)$ is true for all $s$
\item $(\mathcal{M},s) \models \sigma$ iff $\sigma \in \mathcal{L}(s_0)$
\item $(\mathcal{M},s) \models \neg \Phi$ iff $(\mathcal{M},s) \not \models \Phi$
\item $(\mathcal{M},s) \models \Phi_1 \wedge \Phi_2$ iff $(\mathcal{M},s) \models \Phi_1$ and $(\mathcal{M},s) \models \Phi_2$.
\item $(\mathcal{M},s) \models \mathbb{P}^\Gamma_J(\phi)$ iff $\mathcal{M}$, $\mathbb{P}[(\mathcal{M}^{\Gamma},s) \models \phi] \in J$ for some scheduler $\Gamma$.
\end{enumerate}
The semantics of PCTL* path formulas are defined as: 
\begin{enumerate}
\item $(\mathcal{M}, \mathit{S}) \models \Phi$ iff $(\mathcal{M}, s_0) \models \Phi$
\item $(\mathcal{M}, \mathit{S}) \models \mathbf{X} \phi$ iff $(\mathcal{M}, \mathtt{Paths}(s_1)) \models \phi$
\item $(\mathcal{M}, \mathit{S}) \models \phi_1 \mathbf{U} \phi_2$ iff $\exists j \geq 0$ such that $(\mathcal{M}, \mathtt{Paths}(s_j)) \models \phi_2$ and for all $k < j, (\mathcal{M}, \mathtt{Paths}(s_k)) \models \phi_1$. 
\end{enumerate}
\end{definition}

Problem \ref{ProblemOrig} can then be interpreted as finding a defender policy to satisfy $\varphi$ under any adversary policy, and then synthesizing a scheduler so that $(\epsilon, \delta)-$differential privacy holds for states along trajectories generated according to the 
policies in the previous step. 
The scheduler will be related to defining a symmetric relation between states, and ensuring that policies computed at these states are `close' to each other so that subsequent states along the trajectories remain related. 
In this manner, the original objective of maximizing the probability of satisfying $\varphi$ will also not be affected. 

\subsection{Distances between States and Differential Privacy}\label{DistStateDiffPriv}

The eavesdropper observes a sequence of labels, corresponding to labels on states along a trajectory. 
The eavesdropper is also assumed to have access to optimal defender policies at each state synthesized in the previous step. 
Informally, a trajectory will be differentially private if at each state along the trajectory, the adversary will be unable to associate a unique state $s = \mathcal{L}^{-1}(l)$, where $\mathcal{L}^{-1}:2^{\mathcal{AP}} \rightarrow 2^S$. 

%
To capture the relation between the notion of differential privacy in Definition \ref{DiffPrivDefn} and paths in a Markov chain, we use the notion of a \emph{skewed distance} from \cite{chistikov2018bisimilarity}.

\begin{definition}[Skewed Total Variation Distance]\label{SkewedTotVarDistDefn}
For $\alpha \geq 1$, the skewed distance is a quantity $\Delta_\alpha:\mathbb{R}_{\geq 0} \times \mathbb{R}_{\geq 0} \rightarrow \mathbb{R}_{\geq 0}$ such that $\Delta_\alpha(x,y) := \max\{x-\alpha y, y-\alpha x, 0\}$. 
Then, given $s, s' \in S$, the skewed total variation distance is 
\begin{align*}
tv_\alpha (s,s') :=tv_\alpha (\nu_s,\nu_{s'})= \sup_{E \in \mathcal{F}} \Delta_\alpha(\nu_s(E), \nu_{s'}(E))
\end{align*}
\end{definition}
%

This gives the following result \cite{chistikov2018bisimilarity}: 
\begin{proposition}\label{PropDiffPrivSkewedTotVarn}
The labeled MC $\mathcal{M}$ is $(\epsilon, \delta)-$differentially private with respect to $R \subset S \times S$ if for every $s, s' \in S$ such that $(s, s') \in R$, we have $tv_\alpha (s,s') \leq \delta$, where $\alpha = e^\epsilon$.
\end{proposition}

To establish differential privacy, we have to define a notion of equivalence on states of the MDP (or Markov chain) so that they are indistinguishable from each other. 
We use a skewed variant of the Kantorovich distance\footnote{This metric has its origins in optimal transportation theory, but has since been used in several applications in computer science \cite{deng2009kantorovich}.}\cite{kantorovich1942transfer} to translate the distance between MDP states to a distance over probability measures \cite{ferns2004metrics}. 
This is needed since the consequence of taking an action in a state often yields a distribution over the next state, which is why we need to `\emph{lift}' $tv_\alpha$ from a relation over states to one over distributions. 

\begin{definition}[Skewed Kantorovich Metric]\label{SkewedKantDistDefn}
Let $\omega, \omega'$ be probability distributions over states $S$ of an MDP $\mathcal{M}$. 
Then, given a symmetric distance $d :S \times S \rightarrow [0,1]$, the skewed Kantorovich distance between $\omega$ and $\omega'$, $K_\alpha^d (\omega, \omega')$, is the maximum value of a pair of linear programs- the first is given below; the second is got by reversing $\omega$ and $\omega'$:
\begin{align}
\max_{x_s} &\sum_{s \in S} (\omega(s)-\alpha \omega'(s))x_s \label{KantDistPrimal} \\
\text{subject to: }&x_s - \alpha x_{s'} \leq d(s,s'), \text{\quad for all }s, s' \in S \nonumber \\
&0 \leq x_s \leq 1,  \text{\quad \quad for all }s \in S \nonumber
\end{align}
\end{definition}
\begin{lemma}\cite{chistikov2018bisimilarity}  
Let $d(s,s') = \mathbf{I}_{\neq}(s,s')$, where $\mathbf{I}_{\neq}$ takes value $1$ if its arguments are not equal, and $0$ otherwise. 
Then $K_\alpha^d (\omega, \omega') = tv_\alpha(s, s')$. 
\end{lemma}
\subsection{Value Iteration for Differential Privacy}\label{ValIterDiffPriv}


The approach we take is to ensure that differential privacy will hold at each state of the trajectory, starting from the initial state. 
To achieve this, we define a symmetric relation between states in a recursive manner. 
We further stipulate that at related states, optimal defender policies synthesized in the previous step are close to each other, and that subsequent states resulting from these policies be related. 
We formalize this approach and prove relevant results in this section. 

We recall standard terminology from the MDP literature \cite{puterman2014markov}. 
The \emph{value} of a state $s$ under scheduler $\Gamma$ is $V^\Gamma(s) : = \mathbb{E}[\sum_{t=0}^\infty \beta^t r_t|s_0=s,\Gamma]$, where $s_0$ is the initial state, $\beta \in (0,1)$ is a discount factor, and $r_t$ is a \emph{reward} received at time $t$. 
The expectation is taken over the transitions of state induced by $\Gamma$. 
The goal is to determine $\Gamma$ to maximize $V^\Gamma(s)$ for each $s$. 
The optimal value function $V^*$ will then satisfy the \emph{Bellman optimality equations}, given by:
\begin{align}
V^*(s)&=\max_{a \in A} (r^a_s + \beta \sum_{s_1 \in S} \mathbb{P}(s_1|s,a)V^*(s_1)) , \forall s \in S \label{BellOptEqn}
\end{align}
\begin{lemma}\cite{puterman2014markov}
Let $V_{0}(s) = 0$ and $V_{n+1}(s) = \max \limits_{a \in A} (r^a_s + \beta \sum_{s_1 \in S}\mathbb{P}(s_1|s,a)V_n(s_1)$. 
Then, $\{V_n(s) \}_{n \geq 1}$ converges uniformly to $V^*(s)$ for each $s \in S$. 
\end{lemma}
%
%

Since our treatment is related to maximizing the satisfaction probability, we will assume that $r^a_s = 0$ at all states other than accepting states of the product game (where this reward will be set to $1$). 
In this case, the value of a state will be related to the probability of satisfying the LTL goal, and therefore will be in $[0,1]$. 
Consequently, we can set $\beta = 1$. 


Define an operator $F_{tv_\alpha}$ as follows:
\begin{align*} 
&F_{tv_\alpha}(s,s'):= \begin{cases} \max \limits_{a \in A}( tv_\alpha(s,s'))&\mathcal{L}(s) = \mathcal{L}(s')\\1&\mathcal{L}(s) \neq \mathcal{L}(s') \end{cases} 
\end{align*}

Let $F^n_{tv_\alpha}$ denote the composition of $F_{tv_\alpha}$ $n$ times.

\begin{theorem}\label{TheoremDiffinValue}
$\Delta_\alpha(V_n(s) , V_n(s')) \leq F^n_{tv_\alpha} (s,s')$ $\forall s, s' \in S$. 
\end{theorem}

\begin{proof}
From the definition of $F_{tv_\alpha} (s,s')$, the claim clearly holds when $\mathcal{L}(s) \neq \mathcal{L}(s')$. 
Now, let $\mathcal{L}(s) = \mathcal{L}(s')$. 
Since $V_0(s) = 0$ for all $s$, the base case of the induction holds. 
Now, suppose $\Delta_\alpha(V_k(s) - V_k(s')) \leq F^k_{tv_\alpha} (s,s')$ is true for some $k$. 
Then, we have:
\begin{align*}
&V_{k+1}(s)-\alpha V_{k+1}(s')\\& = \max_{a} [\sum_{u \in S} \mathbb{P}(u|s,a)V_n(u)]  - \alpha \max_a [\sum_{u \in S} \mathbb{P}(u|s',a)V_k(u)]\end{align*} \begin{align*}
&\leq \max_a [\sum_{u \in S} (\mathbb{P}(u|s,a)-\alpha \mathbb{P}(u|s',a))V_k(u)]
\end{align*}
Writing down the above set of inequalities for $V_{k+1}(s')-\alpha V_{k+1}(s)$, we observe that $V_k(u)$ is a feasible solution to the skewed Kantorovich metric $tv_\alpha^k(s,s')$ (Definition \ref{SkewedKantDistDefn}). 
Therefore, we get:
$\Delta_\alpha(V_{k+1}(s) - V_{k+1}(s')) \leq \max_a [tv_\alpha^k(s,s')]
=F_{tv_\alpha}(F^k_{tv_\alpha}(s,s')) = F^{k+1}_{tv_\alpha}(s,s')$, completing the proof.
\end{proof}

\begin{proposition}\label{PropnValConv}
Let $F^*_{tv_\alpha}(s,s')$ be the least fix point of $F_{tv_\alpha} (s,s')$. As $n \rightarrow \infty$, $\Delta_\alpha (V^*(s), V^*(s'))$ $\leq F^*_{tv_\alpha}(s,s')$.
\end{proposition}

\begin{proof}
This follows from the fact that in $K_\alpha^d(\omega, \omega')$, if we equip $d: S \times S \rightarrow [0,1]$ with a point-wise ordering, we get a complete lattice. When $K_\alpha^d(\omega, \omega') = tv_\alpha(s,s')$, then $F_{tv_\alpha} (s,s')$ is monotone with respect to this order. 
Consequently, it admits a least fixed point \cite{tarski1955lattice}. 
\end{proof}


These results allow us to aggregate related states. 
Further, we restrict the actions (the set $A$ in Equation (\ref{BellOptEqn})) to those determined by the optimal defender policy $\mu^*$, where $\mu^*:=\arg \max \limits_\mu \min \limits_\tau \mathbb{P}(\varphi)$. If $\mu^*(s)$ denotes the actions available at state $s$ per the policy $\mu^*$, we write 
$A(s):=\{a : a \in \mu^*(s)\}$. 

\begin{assumption}\label{AssSatisTraj}
Assume 
that at each non-terminal state $s$ along the trajectory that satisfies $\varphi$, there is some action $a$ such that the LTL objective will be satisfied.
\end{assumption}

Assumption \ref{AssSatisTraj} stipulates that trajectories that will be generated according to the policies $\mu^*$ in Section \ref{SynthDefPolicy} satisfy the LTL objective with non-zero probability. 
A more thorough analysis of ensuring differential privacy of trajectories that may not satisfy the LTL goal at all is left as future work. 

In the sequel, we abuse notation to say that a state of the labeled MDP is a \emph{terminal state} if it corresponds to an accepting state of the product game formed by composing representations of the environment and the LTL objective. 

\begin{theorem}\label{TheoremSymmReln}
Consider a symmetric relation $R$ defined recursively as: $(s,s') \in R$ if and only if:
\begin{enumerate}
\item $\mathcal{L}(s) = \mathcal{L}(s')$,
\item $tv(\mu^*(s), \mu^*(s')) \leq M$ where $M << 1/\alpha$, 
\item $\exists u, u' \in S, a_s \in \mu^*(s), a_{s'} \in \mu^*(s')$ such that $\mathbb{P}(u|s,a_s) > 0, \mathbb{P}(u'|s',a_{s'}) > 0$, and $(u,u') \in R$ for all non-terminal states $u, u'$, 
\item $tv_{\alpha} (s,s') = 0$ for terminal states $s,s'$. 
\end{enumerate}
Consider two trajectories, one starting from $s_0 \in S$, and the other from $s_0' \in S$. 
Then, if $(s_0,s_0') \in R$, each state along the two trajectories, written $(s,s')$, will be $(\epsilon, \delta)-$differential private for $\alpha = e^\epsilon$ if $\delta \geq \alpha M+  \max \limits_{(s,s') \in R} F^*_{tv_\alpha}(s,s')$. 
\end{theorem}

\begin{proof}
$F^*_{tv_\alpha}(s,s') \geq tv_\alpha (s,s')$ since $F^*_{tv_\alpha}$ is a least fixed-point. 
From Thm. \ref{TheoremDiffinValue} and Proposition \ref{PropnValConv}, $\Delta_\alpha (V^*(s), V^*(s'))$ $ \leq  F^*_{tv_\alpha}(s,s')$. 
The second condition for $R$ allows us to relax the requirement that the optimal action from $s$ and that from $s'$ is the same. 
This will allow two states to be related if the optimal (stochastic) defender policies from these states are `close' to each other. 
The third condition ensures that these policies lead to transitions to states that will also be related to each other. 
Now, since $V^*(s) = \max_\Gamma \mathbb{P}(\varphi|s)$, where $\Gamma$ comprises actions from $\mu^*$, $\Delta_\alpha (V^*(s), V^*(s'))$ is $|\mathbb{P}(\varphi|\mathtt{Paths}(s))-\alpha \mathbb{P}(\varphi|\mathtt{Paths}(s'))|$. 
This is the same as $tv_\alpha(s,s')$, but along measurable sets determined by $\Gamma$. From Proposition \ref{PropDiffPrivSkewedTotVarn}, we want this quantity, plus an additional term that accounts for the closeness of optimal policies to be less than $\delta$. 
Therefore, $\delta \geq  F^*_{tv_\alpha}(s,s') + \alpha M$. 
The recursive definition of $R$ requires that this property be maintained at every subsequent state until a terminal state is reached. 
The lower bound on $\delta$ is got by taking the maximum over least fixed points corresponding to each pair of related states.
\end{proof}
In Theorem \ref{TheoremSymmReln}, we compare trajectories of equal lengths. 
The length of a trajectory is the number of actions in the trajectory, plus one (for the initial state). 
Future work will study differential privacy for trajectories of different lengths.

\begin{remark}
Computing $tv_\alpha (s,s')$ is not known to be decidable \cite{kiefer2018computing}. 
However, in our setting, since $V(s)= \mathbb{P}(\varphi|s)$, Theorems \ref{TheoremDiffinValue}, \ref{TheoremSymmReln}, and Proposition \ref{PropnValConv} will allow us to circumvent the need to explicitly compute $tv_\alpha (\cdot, \cdot)$.
\end{remark}
%
%
	%
\section{Example}\label{Example}

We consider a CPS in the form of a drone/ UAV that has to carry out persistent surveillance of a target region. 
At the same time, it must avoid certain regions of the environment. 
This could be either due to the presence of a physical obstacle, or other factors that might compromise it (e.g. entering a region might allow the drone to be detected by a radar). 
This goal can be represented by the LTL formula $\varphi= \mathbf{G}\mathbf{F} \mathtt{tar} \wedge \mathbf{G} \neg \mathtt{obs}$, where $\mathtt{tar}$ and $\mathtt{obs}$ are labels denoting the target and obstacle respectively. 
The DRA corresponding to $\varphi$ will have two states $q_0, q_1$, with $F = (\{\emptyset\}, \{q_1\})$. 

The dynamics model of the UAV motion is inspired from \cite{kress2007s}. 
The satisfaction of $\varphi$ can be affected by an adversary that can influence trajectories of the UAV. 
A stochastic-game abstraction of the UAV dynamics was presented in \cite{niu2019optimal}, and we will assume that we have this abstraction for the remainder of this section. 
The environment of the drone is then an $M \times N$ grid, $S:=\{s_i:i = x+My, x \in \{0,\dots,M-1\}, y \in \{0, \dots, N-1\}\}$. 
%
We assume that the drone's actions are $U_{def} = \{R, L, U, D\}$ denoting right, left, up, and down, and the actions of the adversary are $U_{adv} = \{A, NA\}$, denoting attack, and not attack respectively. 
Transition probabilities for $(u_{def}, u_{adv}) = (R, NA)$ and $(R, A)$ are defined below. Probabilities for other action pairs can be defined similarly. 
Let $N_{s_i}$ denote the neighbors of $s_i$.

%
%
\[
\mathbb{T}(s_j|s_i, R, NA)=
\begin{cases}
0.8 & \text{$j=i+1$, $(i+1)\not\equiv 0 \bmod M$}\\ 
\frac{0.2}{|N_{s_i}|} & \parbox{0.23 \textwidth}{($s_j \in  \{s_i\} \cup N_{s_i}\setminus \{s_{i+1}\}$), $(i+1) \not\equiv 0 \bmod M$}\\
1 & \text{$j =i$ and $(i+1) \equiv 0 \bmod M$}\\
\end{cases}\]
\[
\mathbb{T}(s_j|s_i, R, A)=
\begin{cases}
0.6 & \text{$j=i+1$, $(i+1)\not\equiv 0 \bmod M$}\\ 
\frac{0.4}{|N_{s_i}|} &\parbox{0.23 \textwidth}{($s_j \in  \{s_i\} \cup N_{s_i}\setminus \{s_{i+1}\}$), $(i+1) \not\equiv 0 \bmod M$}\\
1 & \text{$j =i$ and $i+1  \equiv 0 \bmod M$}
\end{cases}
\]

We use the method from Section \ref{SynthDefPolicy} to synthesize a drone policy that will maximize its probability of satisfying this objective under any adversary policy. 
Once this has been done, we want to ensure that a second, independent adversary eavesdrops on a labeled trajectory that satisfies the LTL objective, will not be able to discern the position of the drone. 
We use the skewed total variation distance in order to establish differential privacy of the trajectory at each state along the trajectory. 
Specifically, two states will be related if they satisfy the conditions in Theorem \ref{TheoremSymmReln}. 

\begin{figure}[!h]
 \centering
  \includegraphics[width=2.25 in]{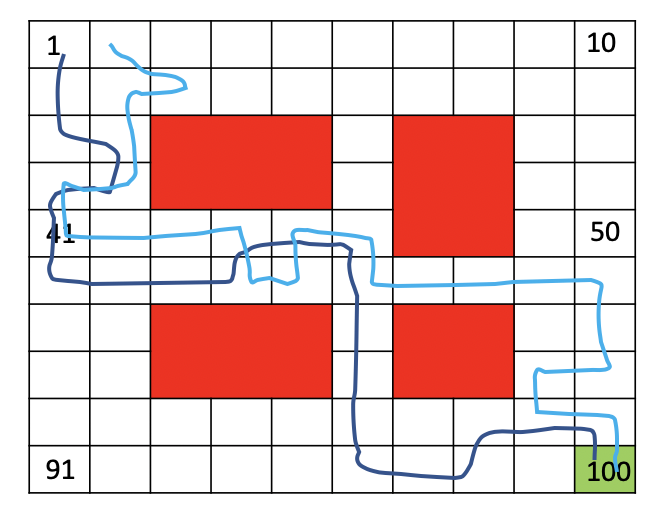} 
\caption{The environment as a $10 \times 10$ grid. The objective for the CPS, given by the LTL formula $\varphi = \mathbf{GF} \mathtt{tar} \wedge \mathbf{G} \neg \mathtt{obs}$, is to visit the target (green state) infinitely often while always avoiding obstacles (red states). The CPS has to satisfy $\varphi$ in the presence of an adversary who can insert inputs that affect transitions between successive states. After a trajectory that satisfies $\varphi$ has been realized, a second eavesdropper adversary should not be to distinguish this trajectory from trajectories `sufficiently close' to it. We use differential privacy as a metric to decide if the latter objective will be satisfied. The figure shows a fragment of length $25$ of two trajectories that satisfy $\varphi$ and are differentially private for $\epsilon = 1, \delta = 0.001$. The trajectories start from states that are related to each other, and at each state, the distance between the optimal defender policies at the states is below a threshold, and subsequent states are related to each other. An eavesdropper will not be able to distinguish between these trajectories only by observing labels on the states.}\label{EnvRep}
\end{figure}
\begin{figure}[!h]
 \centering
  \includegraphics[width=2.75 in]{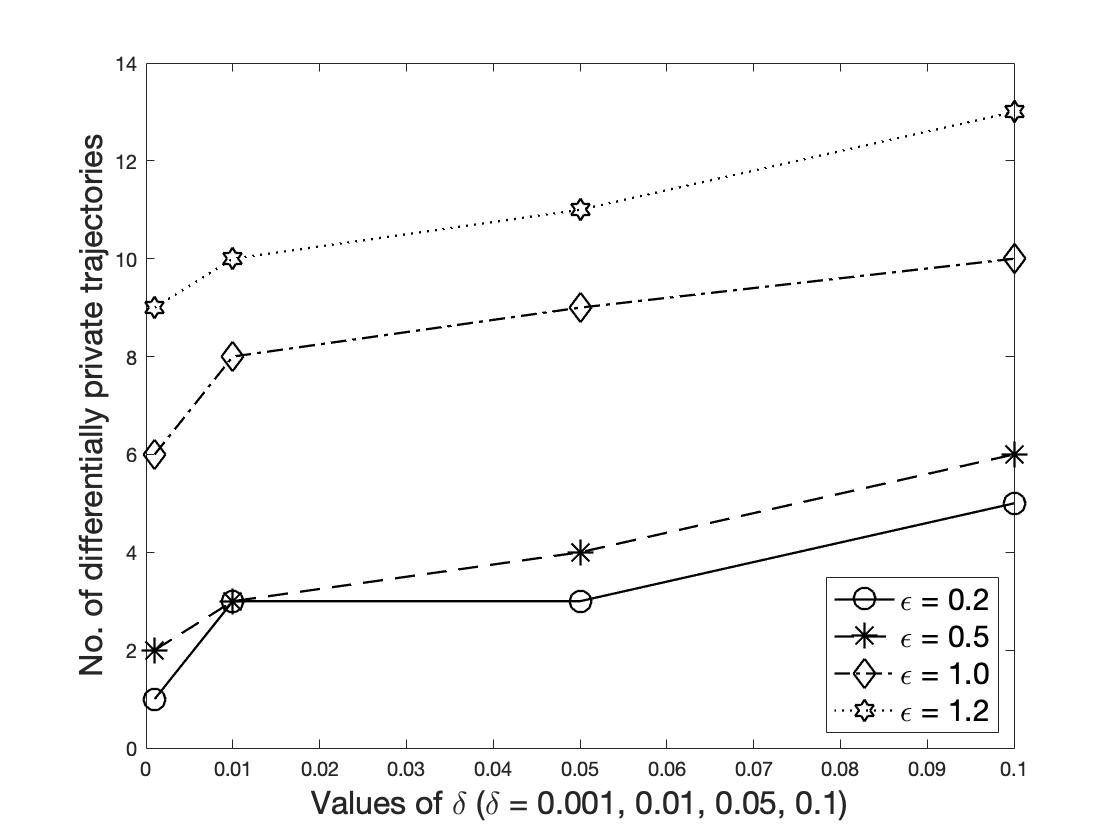} 
\caption{Number of trajectories differentially private to one satisfying trajectory for different values of $\epsilon$ and $\delta$. Larger $\epsilon$ and $\delta$ allow for more differentially private trajectories.}\label{NumDiffPrivTraj}
\end{figure}

We report results of our experiments for $M = N = 10$, as shown in Figure \ref{EnvRep}. 
In the first step, using the results in Section \ref{SynthDefPolicy}, we synthesize a policy for the defender that will maximize the probability of satisfying $\varphi$ under any adversary policy. 
We consider a trajectory that satisfies $\varphi$, and use results in Sections \ref{DistStateDiffPriv} and \ref{ValIterDiffPriv} to observe the effects of values of $\epsilon$ and $\delta$ on the number of differentially private trajectories that can be realized.  
Specifically, different values of $\epsilon$ and $\delta$ will yield varying numbers of differentially private trajectories, under an additional assumption that the optimal defender policies at states which are related to each other are sufficiently close. 
In particular, we observe that larger values of $\epsilon$ and $\delta$ allow for more differentially private trajectories, as seen in Figure \ref{NumDiffPrivTraj}. 
Figure \ref{EnvRep} also shows a fragment of length $25$ of two trajectories that are $(1, 0.001)-$differentially private. 
That is, by simply observing labels on the states, and with knowledge of the optimal defender policies at each state, with high probability, the eavesdropper adversary will not be able to distinguish between the two trajectories. 

	
\section{Conclusion}\label{Conclusion}

This paper presented a solution to the problem of ensuring satisfaction of an LTL objective $\varphi$ while maintaining privacy of trajectories in the presence of two kinds of adversaries. 
We modeled the interaction between the CPS and an adversary who could tamper with actuator inputs as a stochastic game. 
Maximizing the probability of satisfying $\varphi$ under actions of this adversary was equivalent to reaching a Stackelberg equilibrium of this game. 
We then characterized the indistinguishability of trajectories to an eavesdropper and showed a way to ensure differential privacy of states along trajectories that satisfied $\varphi$. 
We showed that if a distance between the probabilities of satisfying of $\varphi$ starting from states that were related to each other was below a threshold, then this ensured that these states were differentially private. 
An additional requirement on the distance between optimal policies at related states ensured differential privacy at every state along the trajectories. 
We validated our approach on a simulation of a UAV that had to satisfy an LTL objective in the presence of adversarial inputs in an environment with an eavesdropper.

Future directions of research include generalizing our setup to the goal of ensuring differential privacy of trajectories that may not satisfy $\varphi$. 

\bibliographystyle{IEEEtran}
\bibliography{DiffPrivTLRef}
\end{document}